\documentclass[a4paper, 12pt]{article}

\usepackage[margin=2cm]{geometry} 

\usepackage{amsmath}
\usepackage{amsthm}
\usepackage{amssymb}
\usepackage{booktabs}
\usepackage{multirow}
\usepackage[utf8]{inputenc}
\usepackage{hyperref}
\hypersetup{
	unicode,
	pdfauthor={Author One, Author Two, Author Three},
	pdftitle={A simple article template},
	pdfsubject={A simple article template},
	pdfkeywords={article, template, simple},
	pdfproducer={LaTeX},
	pdfcreator={pdflatex}
}

\usepackage{natbib}
\theoremstyle{plain}

\theoremstyle{definition}

\usepackage{url}

\usepackage{graphicx, color}
\graphicspath{{fig/}}

\usepackage{algorithm, algpseudocode} 
\usepackage{mathrsfs} 

\usepackage{lipsum}

\title{\emph{Bayes-xG}: Player and Position Correction on Expected Goals (xG) using Bayesian Hierarchical Approach}
\author{Alexander Scholtes$^1$ \and Oktay Karaku\c{s}$^1$}

\date{
	$^1$Cardiff University, School of Computer Science and Informatics, Abacws, Senghennydd Road, Cardiff, CF24 4AG, UK.\\ \texttt{\{scholtesa, karakuso\}@cardiff.ac.uk}\\[2ex]%
}
\usepackage{lineno}
\usepackage{setspace}
\begin{document}
\maketitle
\begin{abstract}
    This study employs Bayesian methodologies to explore the influence of player or positional factors in predicting the probability of a shot resulting in a goal, measured by the expected goals (xG) metric. Utilising publicly available data from StatsBomb, Bayesian hierarchical logistic regressions are constructed, analysing approximately 10,000 shots from the English Premier League to ascertain whether positional or player-level effects impact xG. The findings reveal positional effects in a basic model that includes only distance to goal and shot angle as predictors, highlighting that strikers and attacking midfielders exhibit a higher likelihood of scoring. However, these effects diminish when more informative predictors are introduced. Nevertheless, even with additional predictors, player-level effects persist, indicating that certain players possess notable positive or negative xG adjustments, influencing their likelihood of scoring a given chance. The study extends its analysis to data from Spain's La Liga and Germany's Bundesliga, yielding comparable results. Additionally, the paper assesses the impact of prior distribution choices on outcomes, concluding that the priors employed in the models provide sound results but could be refined to enhance sampling efficiency for constructing more complex and extensive models feasibly.
    
\end{abstract}


\section{Introduction}
\label{sec:intro}
One of the most common advanced football analytics metrics is the idea of expected goals (xG), which estimates the probability of a given shot resulting in a goal based on several features about the shot such as distance from the shooter to the goal or the body part used by the shooter. However, none of the mainstream xG models take into account any player-specific features when estimating these values. To illustrate this, imagine that you have two players taking the same shot from the same position, with defenders in the same place and everything else being the same. Still, one player is Lionel Messi and the other is a random player from the National League (English 5th tier). Obviously, players who play in the National League are good, but it is not unreasonable to assume that Lionel Messi would be more likely to score. However, xG metrics would assign the same value for both of these changes.

The objective of this paper is to investigate if there are position or player effects on xG, meaning that certain positions or players have higher or lower goal probabilities for a given chance than others. This will be achieved using a Bayesian hierarchical model, where the hierarchies will be the position of the player or the player. The results of this method will initially be compared to a more traditional frequentist xG model to evaluate baseline results without any group effects. Then the hierarchical models will be compared to the non-hierarchical Bayesian models, to assess the impact of having hierarchies in the data on the results. If the rationale described above of the two players taking the same shot is valid, then it is expected that the xG predictions of the hierarchical models will differ significantly from the non-hierarchical models, supporting the idea that there is a position and/or player effect on the xG of a given shot.

This paper will begin with a review of the relevant literature around this topic, looking at the development of football analytics and xG, as well as any attempts to use Bayesian modelling in football analytics. Thereafter, the methodology will be described by going through the frequentist and Bayesian techniques used. Then, the data will be introduced and described with any changes made before the choice of Bayesian prior distributions for predictors is discussed. Next, the results of the modelling will be presented before validating the results of the Bayesian models on additional data. The aforementioned choice of prior distributions will then be evaluated. Finally, a discussion section will deliberate on the significance of the results before concluding the paper.

\section{Related Works}
\label{sec:litRev}
The use of data in football is often not fully embraced, with many decision-makers arguing the sport is too complex for data to be used effectively to improve results and performance \citep{smith2022expected}. However, with its successful use in other sports, there was sufficient interest for some clubs, companies, and individuals to pursue using data to derive conclusions and make suggestions in football. With a growing demand for data, companies that specialise in sports data collection have grown too, along with their ability to track data. The result is that there is now an enormous amount of football data to use for several purposes, such as player/club performance, scouting, and player fitness and injury risk, to name a few \citep{tippett2019expected}.

At the heart of the idea of using data in football was the potential to gain a competitive advantage. As a result, clubs that use data tend to be secretive about their operations and procedures \citep{tippett2019expected}. Despite this, there is plenty of publicly available literature and sources showing how data can be used in football. Moreover, sports broadcasters have long used data when giving an overview of a match, such as possession statistics. Still, these have only recently moved away from simple counts and percentages to more complex metrics. The Bundesliga, for example, provides a goal probability value after each goal is scored, giving the chance of that given opportunity resulting in a goal \citep{Aberle_Figdor_Mongrand_Janetzke_2020}.

This goal probability, also commonly called expected goals (xG), has been a central topic in the development of more advanced statistics using football data \citep{smith2022expected}. Crucially, it moves away from the idea of things that did happen and focuses on things that could have happened. With football being such a complex and chaotic sport, outcomes often do not reflect expectations as matches are often decided by fine margins or decisions out of the players’ and coaches’ control. Nevertheless, using expectations gives decision-makers an idea of the underlying performance of their team and allows them to see if their team is over or underperforming according to expectations \citep{brechot2020dealing}.

There have been many versions of xG models created since the idea was founded, using a variety of machine-learning techniques and data sources. \cite{herold2019machine} provide a summary of many applications of machine learning in football, including xG models. The most common methods of estimating xG in their paper are logistic regressions, decision trees, ensemble methods (e.g. random forest), and neural networks. \cite{lucey2015quality} use player and ball tracking data from the 10 seconds leading up to a shot to estimate goal probabilities across an entire season and found that “defender proximity, interaction of surrounding players, speed of play, coupled with shot location play an impact on determining the likelihood of a team scoring a goal”. \cite{madrero2020creating} uses qualitative data from the popular video game FIFA to account for player effects on xG using a logistic regression and an XGBoost model. They found that an adjusted model can better predict goals over a season for individual players and teams than an overall xG model. \cite{fairchild2018spatial} built an xG model again using logistic regression and used it to estimate MLS teams’ offensive efficiency in scoring. They also discuss evaluation metrics for expected goals models and suggest the use of the Brier score to compare predicted probability to ground-truth binary outcomes. \cite{cavus2022explainable} apply a variety of ensemble and boosting methods to calculate xG values and find that a random forest model performs best, even compared to models from other papers using other techniques and data.

The closest study to this paper to date is that of \cite{hewitt2023machine}, which investigates position and player-adjusted xG models. They find evidence of positional adjustments with forwards having a positive adjustment, midfielders having a slightly negative adjustment, and defenders having a large negative adjustment. Moreover, they also find evidence of player effects on xG by fitting their model with only data from Lionel Messi and find a large positive adjustment in this case.

One of the features a lot of these models have in common is their frequentist approach, as opposed to using Bayesian methods. \cite{spearman2018beyond} uses a Bayesian approach to estimate the maximum a \emph{posteriori} effects of parameters in a model for predicting future scoring of teams in games. \cite{joseph2006predicting} used Bayesian networks and Naïve Bayes learners to predict the results of matches played by Tottenham Hotspur and compared the results to K-nearest neighbour and decision tree models. They reiterate one of the benefits of Bayesian modelling which is comparably accurate predictions in the absence of a large amount of data. \cite{zambom2018determinants} use Bayesian methods to analyse team performance in Europe’s top leagues to determine which features tend to be most significant in predicting team performance. They find that the most important features include the number of assists, the number of shots conceded, saves made by the goalkeeper, passing accuracy, and number of shots on target.

One area of Bayesian modelling which is also often not considered in football analytics is using multi-level, or hierarchical, models. \cite{tureen2022estimated} construct a multi-level model for player-adjusted expected goals but do not use a Bayesian approach to do so. Still, they use their model to calculate estimated player impact values on xG. On the other hand, \cite{baio2010bayesian} construct a Bayesian hierarchical model but use it to predict match results as opposed to xG directly and group their data by the team as opposed to by the player. Still, the use of Bayesian hierarchical modelling in football is a relatively unexplored area.

By using Bayesian hierarchical modelling, group-level effects can be reliably estimated even with small group sizes. Therefore, the effect of a player’s position or even the player themselves on the chance of a given shot resulting in a goal can be reliably measured. The result is that certain players could be identified as being more likely to score than others for given chances, which is a result that can be used for player selection or scouting purposes. This idea is present in the work of \cite{hewitt2023machine}, where Messi is found to be an extremely efficient goal scorer. This conclusion may be obvious to football fans, but the fact that the efficiency can be reliably measured is extremely interesting for potentially comparing the goalscoring efficiency of footballers. \cite{tureen2022estimated} construct a metric they refer to as “estimated player impact”, which is another calculation of a player’s individual effect on the probability of scoring. The estimation of a player’s impact on xG can potentially be another tool in evaluating player performance for team selection or scouting purposes.

\section{Methodology}
\label{sec:method}
After a general look at the literature on the xG metric and its evaluation throughout the years, in the section, we explain the details of the methodology this paper is proposing to study positional and player-related corrections to generic xG approaches. We propose the utilisation of the Bayes formula by taking player and position information into a conditional probability formulation which is then evaluated under Bayesian hierarchical modelling. 

Before showing how Bayesian methods can be applied to xG modelling, we now show how xG models are typically created. This involves using a frequentist approach and, in this case, a logistic regression appears as the natural choice to obtain goal probabilities. This paper will first attempt to build a generic xG model with comparable results to an xG model built by StatsBomb, an industry leader in data collection and analysis. To do so, we gradually increase the number of predictors (both given and engineered) used in the logistic regression that generally follows the formulation given below

\begin{align}\label{equ:logistic}
    logit(p_i)=\log\left(\dfrac{p_i}{1-p_i}\right)=\beta_0+\sum_{j=0}^N\beta_j\cdot X_{ji}
\end{align}
where $p_i$ is the probability of the shot $i$ resulting in a goal, $X_{ji}$ is the value of predictor $j$ for the shot $i$. Traditional logistic regression can be seen as a method used to model the relationship between a binary dependent variable (in this context Goal or No-Goal) and one or more independent variables (given and engineered features). It uses the logistic function in (\ref{equm:logistic}) to transform a linear combination of features into a probability of the dependent variable being one of the two classes.

\subsection{Bayesian Logistic Regression}
Bayesian logistic regression extends logistic regression by introducing a Bayesian framework for modelling. Instead of estimating fixed model parameters as in traditional logistic regression, Bayesian logistic regression treats these parameters as random variables with associated probability distributions. This means that we not only get point estimates for the model parameters but also full probability distributions, allowing us to quantify uncertainty.

In Bayesian logistic regression, we specify prior distributions for the model parameters, representing our beliefs about their values before observing any data. The likelihood function represents the probability of observing the data given the model parameters. Using Bayes' theorem, we update the prior distribution with the likelihood distribution to obtain the posterior distribution, which reflects our updated beliefs after considering the data. To compute the posterior distribution in Bayesian logistic regression, various techniques can be used, including Markov Chain Monte Carlo (MCMC) methods and variational inference. These methods sample from the posterior distribution of the parameters to estimate their values and uncertainties.

Once we have the posterior distribution of the model parameters, one can perform various tasks such as parameter estimation, uncertainty quantification, and prediction. Bayesian logistic regression provides a natural way to make probabilistic predictions, as it generates a distribution of predicted probabilities for each class, rather than just point estimates.

For this paper, the Bayesian methods used will involve fitting Bayesian logistic regressions as both single-level models (baseline) without group-level effects, and multi-level, or hierarchical, models with group-level effects. The predictions from both models will then be compared to determine if there is evidence of group-level effects on xG. To formulate such Bayesian models, this paper specifies
\begin{align}
     &Y_i = \text{binary outcome of shot } i \quad (1 = \text{goal}, 0 = \text{no goal}),\\
     &p_i = \text{probability of shot } i \text{ resulting in a goal},\\
     &X_{n,i} = \text{the value of predictor } n \text{ for shot } i,
\end{align}
where the likelihood distribution is
\begin{align}
    Y_i \sim Bernoulli(p_i)
\end{align}
Hence, the Baseline and Hierarchical models are like
\begin{align}
    &\text{Baseline Model:}\quad logit(p_i)=\beta_0+\sum_{j=0}^N\beta_j\cdot X_{ji},\\
    &\text{Hierarchical Model:}\quad logit(p_{ij})=\beta_0+\sum_{j=0}^N\beta_j\cdot X_{ji} + \beta_{N+1}\cdot X_{N+1,i},
\end{align}
where $N+1$ is the index of the grouping variable in the data.

\subsection{Data}\label{sec:data}
The data used for this project is all freely available event data from StatsBomb, obtained using their Python package \emph{StatsBombPy} (Please see the Statsbomb GitHub page via \url{https://github.com/statsbomb} for details). From their database, only men’s competitions were used because it could be that there is a difference in given goal probabilities in men’s and women’s football, and we have more data from men’s competitions. Then, all open-play shots were extracted with all relevant information for each shot. Set-pieces were excluded because again goal probabilities could vary for set-pieces, and we are not interested in modelling this effect.

The resulting data has more than 60,000 shots from a variety of competitions and years, with 42 columns of information for each shot. Tables \ref{tab:stats1} and \ref{tab:stats2} give some summary statistics for the most relevant variables in the data.

\begin{table}[ht!]
  \centering
  \caption{Summary statistics for most relevant features for xG in the dataset. \textit{N($x$):$y$ means that for $y$ number of samples the variable has the value of $x$.}}\small
    \begin{tabular}{p{4cm}p{7cm}p{2.5cm}p{2.5cm}}
    \toprule
    Variable & Description & \multicolumn{2}{p{5cm}}{Summary (2.d.p)} \\
    \midrule
    \multirow{4}[2]{*}{\textit{distance\_to\_goal}} & \multirow{4}[2]{*}{\parbox{7cm}{Distance between the shooter and the middle of the goal line, normalised to StatsBomb pitch size of 120x80 due to varying pitch sizes.}} & N: 63,309 & Min: 0.63 \\
       &    & Mean: 18.96 & Max: 88.83 \\
       &    & SD: 8.58 & \multicolumn{1}{r}{} \\
       &&&\\
    \midrule
    \multirow{3}[2]{*}{\textit{shot\_angle}} & \multirow{3}[2]{*}{\parbox{7cm}{By creating a triangle between the shooter and the two goalposts, the shot angle is the angle that is by the shooter.}} & N: 63,309 & Min: 0.66 \\
       &    & Mean: 25.39 & Max: 168.61 \\
       &    & SD: 15.60 & \multicolumn{1}{r}{} \\
    \midrule
    \multirow{3}[2]{*}{\textit{gk\_distance\_to\_goal}} & \multirow{3}[2]{*}{\parbox{7cm}{Same as distance\_to\_goal but for the goalkeeper instead of the shooter.}} & N: 63,309 & Min: 0.00 \\
       &    & Mean: 3.56 & Max: 118.00 \\
       &    & SD: 2.61 & \multicolumn{1}{r}{} \\
    \midrule
    \multirow{6}[2]{*}{\textit{players\_in\_shot\_triangle}} &\multirow{6}[2]{*}{\parbox{7cm}{The number of players in the shot triangle, created by the shooter and the two goalposts.}} & N(0): 1,786 & N(1): 30,262 \\
       &    & N(2): 19,481 & N(3): 6,802 \\
       &    & N(4): 2,918 & N(5): 1,217 \\
       &    & N(6): 513 & N(7): 211 \\
       &    & N(8): 77 & N(9): 29 \\
       &    & N(10): 11 & N(11): 2 \\
    \midrule
    \multirow{3}[2]{*}{\textit{opponents\_in\_radius}} & \multirow{3}[2]{*}{\parbox{7cm}{The number of opposition players in a 1m radius of the shooter.}}& N(0): 55,536 & N(1): 7,030 \\
       &    & N(2): 662 & N(3): 71 \\
       &    & N(4): 10 & \multicolumn{1}{r}{} \\
    \midrule
    \multirow{2}[2]{*}{\textit{shot\_body\_part}} & \multirow{2}[2]{*}{\parbox{7cm}{The body part used by the shooter to hit the ball.}} & N(Preferred Foot): 30,738 & N(Head): 10,647 \\
       &    & N(Other Foot): 11,733 & N(Other): 191 \\
    \midrule
    \textit{shot\_first\_time} & Whether the shot was a first-time shot, meaning the shooter took no additional touches of the ball before shooting. & N(True): 20,946 & N(False): 42,363 \\
    \midrule
    \textit{gk\_in\_shot\_triangle} & Whether the goalkeeper was in the shot triangle created by the shooter and the two goalposts when the shot was taken. & N(True): 60,570 & N(False): 2,739 \\
    \midrule
    \textit{shot\_one\_on\_one} & Whether the shooter was one-on-one with the goalkeeper when shooting.& N(True): 3,546 & N(False): 59,763 \\
    \midrule
    \textit{shot\_open\_goal} & Whether the shooter was shooting at an open goal. & N(True): 736 & N(False): 62,573 \\
    \bottomrule
    \end{tabular}%
  \label{tab:stats1}%
\end{table}%

\begin{table}[ht!]
  \centering
  \caption{(\textbf{CON'T}) Summary statistics for most relevant features for xG in the dataset. \textit{N($x$):$y$ means that for $y$ number of samples the variable has the value of $x$.}}\small
    \begin{tabular}{p{3cm}p{7cm}p{3cm}p{3cm}}
    \toprule
    Variable & Description & \multicolumn{2}{p{6cm}}{Summary (2.d.p)} \\
    \midrule
    \multirow{7}[2]{*}{\textit{shot\_technique}}& \multirow{7}[2]{*}{\parbox{7cm}{The technique the shooter used.}} & \multicolumn{2}{p{6cm}}{N(Normal): 47,854}\\ &&\multicolumn{2}{p{6cm}}{N(Overhead Kick): 385} \\
       &    & \multicolumn{2}{p{6cm}}{N(Half Volley): 9,371}\\ && \multicolumn{2}{p{6cm}}{N(Diving Header): 284} \\
       &    & \multicolumn{2}{p{6cm}}{N(Volley): 4,483}\\ && \multicolumn{2}{p{6cm}}{N(Backheel): 244} \\
       &    & \multicolumn{2}{p{6cm}}{N(Lob): 688}\\ 
       \midrule
    \textit{under\_pressure} & Whether the shooter was under pressure when shooting. & N(True): 16,149 & N(False): 47,160 \\
    \midrule
    \textit{goal} & Whether the shot resulted in a goal. & N(True): 6,559 & N(False): 56,750 \\
    \midrule
    \multirow{3}[2]{*}{\textit{shot\_statsbomb\_xg}} & \multirow{3}[2]{*}{\parbox{7cm}{StatsBomb’s own estimated xG value for each shot.}} & N: 63,309 & Min: 0.00 \\
       &    & Mean: 0.10 & Max: 1.00  \\
       &    & SD: 0.13 &  \\
    \midrule
\multirow{3}[2]{*}{\textit{general\_position}} & \multirow{3}[2]{*}{\parbox{7cm}{The general position of the shooter (striker, attacking midfielder, other midfielder, or defender).}} & N(ST): 17,073 & N(M): 15,858 \\
       &    & N(AM): 20,065 & N(D): 10,313 \\
       &&&\\
    \midrule
    \multirow{4}[2]{*}{\textit{player}} & \multirow{4}[2]{*}{\parbox{7cm}{The name of the shooter.}} & N(Messi): 1,907 & N(T. Henry): 325 \\
       &    & N(L. Suárez): 596 & … \\
       &    & N(Iniesta): 374 & N(V. Migas): 1 \\
       &    & N(Neymar): 352 & N(R. Tricella): 1 \\
    \bottomrule
    \end{tabular}%
  \label{tab:stats2}%
\end{table}%

As well as the information given in the columns already in the data, there are several features which were not included by StatsBomb which could be useful for predicting goal probability. Many sources cite distance to goal and shot angle to be two of the most important predictors of goal probability.

The data has the location of the shot, and the StatsBomb data specification \citep{Statsbomb} provides information about the coordinates of the goalposts. Distance to goal is therefore calculated as the Euclidean distance from the shot to the centre of the goal.

For shot angle, the cosine rule is used to calculate the angle from the shooter to the two goalposts. To calculate this reliably, any shots that are taken from the same x-coordinate as the goal line are excluded since this would create a straight line instead of a triangle with no shot angle.

Next, the freeze-frame feature in the data is utilised, which provides information about all other players than the shooter when the shot is taken, including crucially their location on the pitch. From this information, several features are added: the goalkeeper's distance to the goal, whether the goalkeeper is present in the shot triangle formed by the shot and two goalposts, the number of players present in the shot triangle, and the number of opponents within a 1m radius of the shooter. Opponents are used for the 1m radius as opposed to all players because the only time a non-opponent in the radius of a shooter would impact goal probability is when they are in the shot triangle, which is already being accounted for. Otherwise, only opponents will try to put pressure or tackle the shooter outside of the shot triangle.

Then, each player’s mode general position is added. These positions are grouped into: strikers, attacking midfielders, non-attacking midfielders, and defenders, and we expect goal probability to fall on average for each group respectively. Wingers are included as attacking midfielders, while wide midfielders and central midfielders are non-attacking midfielders. Wing-backs are classed as defenders, and so are the few attempts by goalkeepers. Otherwise, all positions to classes are self-explanatory.

Finally, we modify the body part used by changing “left foot” and “right foot” to “preferred foot” and “other foot” according to the player’s apparent preference. To assign these, we go back through the open data and instead look at the passes for each player, we then assign whichever foot the player made the most passes with as their preferred foot because they often have more time to choose which foot to pass with and will then tend to go safely with their preferred foot, which is less feasible when taking a shot as players tend to be under more pressure and have less time.

\subsection{Reference Models}
As outlined in the preceding sections, the primary aims of this paper are to support the adoption of the Bayesian hierarchical modelling approach in investigating the xG metric, specifically concerning player-specific attributes and position groups of players. To fulfil these objectives, we commence by providing definitions for the benchmark and baseline models in the subsequent sections.

\textbf{Statsbomb xG Model:} This model, employed for comparison in this study, relies on Statsbombpy's open data set to provide corresponding xG calculations for each shot discussed in the dataset. The specifics of Statsbomb's xG model remain undisclosed, and it serves as the benchmark model in this paper.

\textbf{Baseline xG Model:} The initial frequentist model introduced is a fundamental model primarily utilising the distance between the shooter and the goal as a predictor, recognised as a key factor in goal probability determination. Another crucial feature considered is the angle of the shot, formed by the shooter and the two goalposts. Recognising the inherent link between distance and shot angle, an interaction term is incorporated to capture the combined influence of both features. This foundational parameterisation constitutes the formulation of the Baseline xG model under a logistic regression model as
\begin{align}
    logit(p_i) = \beta_0 + \beta_1 \cdot (\text{distance\_to\_goal})_i + \beta_2 \cdot (\text{shot\_angle})_i + \beta_3 \cdot (\text{distance}_i \cdot \text{angle}_i)
\end{align}

\begin{figure}[ht!]
    \centering
    \includegraphics[width=0.99\linewidth]{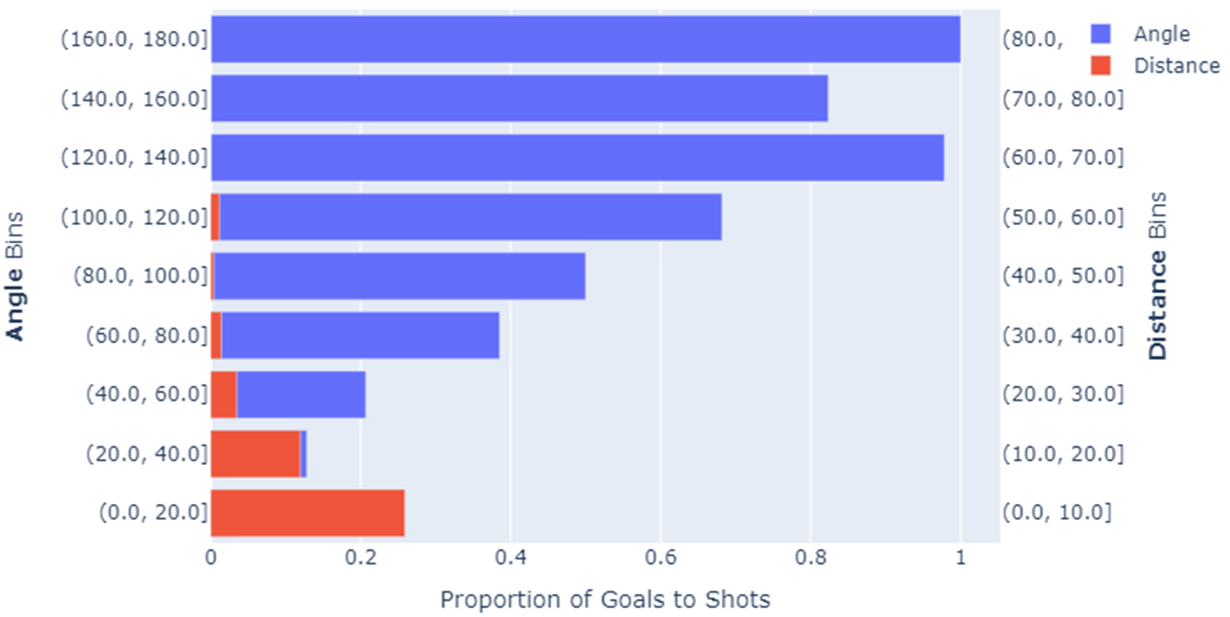}
    \caption{Relationships between shot angle (binned in the 20s)/distance to goal (binned in 10s) and the proportion of goals from shots.}
    \label{fig:disang}
\end{figure}

It is logical to infer that, under reasonable assumptions, the likelihood of scoring decreases, on average, as the distance to the goal increases. This implies that the coefficient $\beta_1$ is likely to be negative. Conversely, $\beta_2$ is expected to be positive, reflecting the observation that as the shot angle decreases, the shot's position is likely to be farther away or from a wider position, both scenarios leading to a lower goal probability. Figure \ref{fig:disang} illustrates the associations between shot angle, distance, and the proportion of goals scored to shots, providing a clearer insight into these relationships.

\textbf{Expanded Model:} While the primary factors considered for calculating the goal probability are typically assumed to be the distance and angle of the shot utilised in the Baseline Model, it is important to acknowledge that various other elements can influence the shot's outcome as shown clearly in the literature. Section \ref{sec:data} has previously addressed each of these factors, and we now introduce our frequentist Extended model, outlined below
\begin{equation} \label{eq:extended}
\begin{split}
    logit(p_i) = \beta_0 + &\beta_1 \cdot \text{distance\_to\_goal}_i + \beta_2 \cdot \text{shot\_angle}_i + \beta_3 \cdot (\text{distance}_i \cdot \text{angle}_i) +\\ &\beta_4 \cdot \text{gk\_distance\_to\_goal}_i + \beta_5 \cdot \text{players\_in\_shot\_triangle}_i +\\ &\beta_6 \cdot \text{body\_part}_i + \beta_7 \cdot \text{first\_time\_shot}_i + \beta_8 \cdot \text{gk\_in\_shot\_triangle}_i +\\ &\beta_9 \cdot \text{one\_on\_one\_shot}_i + \beta_{10} \cdot \text{open\_goal}_i +\beta_{11} \cdot \text{technique}_i +\\ &\beta_{12} \cdot \text{under\_pressure}_i
\end{split}
\end{equation}

\subsection{\emph{Bayes-xG} Models}
Bayesian logistic regression hierarchical modelling is a powerful statistical approach used to analyse and model complex relationships within data. In this context, an additional grouping parameter is employed to incorporate the hierarchical structure of the data, acknowledging potential dependencies or variations across groups. Unlike frequentist logistic regression, Bayesian hierarchical modelling allows for the inclusion of prior information, facilitating a more robust estimation of parameters and uncertainties. Below, we initiate by formulating our \emph{Bayes-xG} models to account for player and position grouping effects. Subsequently, we provide a comprehensive explanation of the rationale behind selecting specific prior distributions for the model parameters.

\subsubsection{Model Definitions}
The foundational Bayesian logistic regression model extends the traditional logistic regression equation by adding a grouping effect
\begin{align}
logit(p_{ik}) = \beta_{0k} + \sum_{j=0}^N\beta_{jk}\cdot X_{ji} + \underbrace{\beta_{N+1,k}\cdot X_{N+1,i}}_{\text{Grouping effect}},
\end{align}
where $k = 1, \dots, \mathcal{K}$ with $\mathcal{K}$ is the number of elements in the group. It is also clear that for this formulation, we extended logistic regression model coefficients $\beta_j \in \mathcal{R}^N$ into a complex form for grouping effect as $\beta_{jk} \in \mathcal{R}^{\mathcal{K}\times (N+1)}$.

Specifically, $\beta_{0k}$ is a \textit{group-specific intercept} for $k^{\text{th}}$ element of the group which accounts for variations in the baseline success probability across different groups. On the other hand, $\beta_{\{1,\dots,N\}k}$ refer to the group-specific slopes which capture variations in the effect of the covariate across groups.

Following the technical details above, for this paper, we decided to define three versions of \emph{Bayes-xG} models which can be expressed as
\begin{itemize}
    \item \emph{Bayes-xG}$_1$ $\rightarrow$ uses the Baseline model with grouping parameter of \textit{position},
    \item \emph{Bayes-xG}$_2$ $\rightarrow$ uses the Extended model with grouping parameter of \textit{position},
    \item \emph{Bayes-xG}$_3$ $\rightarrow$ uses the Extended model with grouping parameter of \textit{player}.
\end{itemize}

\subsubsection{Choice of priors}
When using Bayesian modelling methods, one important consideration is the choice of prior distributions for the predictors in the model. Table \ref{tab:priors} lists the predictors in each of the models and the prior distributions used for them.

\begin{table}[ht!]
  \centering
  \caption{Listing which predictors are used in each of the Bayesian models, and what prior distribution is given to the coefficient of the given predictor.\textit{( $\mathcal{N}$: Normal distribution, $\mathcal{S}\mathcal{N}$: Skew-Normal distribution, $\mathcal{H}\mathcal{N}$: Half-Normal distribution.)}}
    \begin{tabular}{p{5cm}p{0.5cm}p{0.5cm}p{0.5cm}p{8.25cm}}
    \toprule
    \multicolumn{1}{r}{} & \multicolumn{3}{c}{\emph{Bayes-xG}} &  \\\cline{2-4}
    Predictor & 1  & 2  & 3 & \multicolumn{1}{l}{Prior} \\
    \midrule
    \textit{Intercept} & \multicolumn{1}{p{.89em}}{\checkmark} & \multicolumn{1}{p{.89em}}{\checkmark} & \checkmark  &  $\mathcal{N}\left(\mu=0,\sigma=5\right)$\\
    \midrule
    \textit{distance\_to\_goal} & \multicolumn{1}{p{.89em}}{\checkmark} & \multicolumn{1}{p{.89em}}{\checkmark} & \checkmark  &  $\mathcal{S}\mathcal{N}\left(\mu=-1,\sigma=5, \alpha = -1\right)$\\
    \midrule
    \textit{shot\_angle} & \multicolumn{1}{p{.89em}}{\checkmark} & \multicolumn{1}{p{.89em}}{\checkmark} & \checkmark  &  $\mathcal{S}\mathcal{N}\left(\mu=1,\sigma=5, \alpha = 1\right)$\\
    \midrule
    \textit{distance\_angle\_interaction} & \multicolumn{1}{p{.89em}}{\checkmark} & \multicolumn{1}{p{.89em}}{\checkmark} & \checkmark  &  $\mathcal{N}\left(\mu=0,\sigma=5\right)$\\
    \midrule
    \textit{gk\_distance\_to\_goal} &    & \multicolumn{1}{p{.89em}}{\checkmark} & \checkmark  & $\mathcal{N}\left(\mu=0,\sigma=5\right)$ \\
    \midrule
    \textit{players\_in\_shot\_triangle} &    & \multicolumn{1}{p{.89em}}{\checkmark} & \checkmark  &  $\mathcal{S}\mathcal{N}\left(\mu=-1,\sigma=5, \alpha = \{5, 4, \dots, -5\}\right)$ where $\alpha$ is determined by the value of this feature (0 players = 5, 1 player = 4, etc.)\\
    \midrule
    \textit{opponents\_in\_radius} &    & \multicolumn{1}{p{.89em}}{\checkmark} & \checkmark  &  $\mathcal{S}\mathcal{N}\left(\mu=-1,\sigma=5, \alpha = \{1, 0, \dots, -2\}\right)$ where $\alpha$ is determined by the value of this feature (0 players = 1, 1 player = 0, etc.)\\
    \midrule
    \textit{shot\_body\_part} &    & \multicolumn{1}{p{.89em}}{\checkmark} & \checkmark  &  $\mathcal{N}\left(\mu=0,\sigma=5\right)$\\
    \midrule
    \textit{shot\_first\_time} &    & \multicolumn{1}{p{.89em}}{\checkmark} & \checkmark  &  $\mathcal{N}\left(\mu=0,\sigma=5\right)$\\
    \midrule
    \textit{gk\_in\_shot\_triangle} &    & \multicolumn{1}{p{.89em}}{\checkmark} & \checkmark  &  $\mathcal{S}\mathcal{N}\left(\mu=0,\sigma=5, \alpha = -2\right)$\\
    \midrule
    \textit{shot\_one\_on\_one} &    & \multicolumn{1}{p{.89em}}{\checkmark} & \checkmark  &  $\mathcal{S}\mathcal{N}\left(\mu=0,\sigma=5, \alpha = 2\right)$\\
    \midrule
    \textit{shot\_open\_goal} &    & \multicolumn{1}{p{.89em}}{\checkmark} & \checkmark  &  $\mathcal{S}\mathcal{N}\left(\mu=0,\sigma=5, \alpha = 4\right)$\\
    \midrule
    \textit{shot\_technique} &    & \multicolumn{1}{p{.89em}}{\checkmark} & \checkmark  &  $\mathcal{N}\left(\mu=0,\sigma=5\right)$\\
    \midrule
    \textit{under\_pressure} &    & \multicolumn{1}{p{.89em}}{\checkmark} & \checkmark  &  $\mathcal{S}\mathcal{N}\left(\mu=0,\sigma=5, \alpha = -2\right)$\\
    \midrule
    \textbf{general\_position} & \multicolumn{1}{p{.89em}}{\checkmark} & \multicolumn{1}{p{.89em}}{\checkmark} & \multicolumn{1}{r}{} &  $\mathcal{S}\mathcal{N}\left(\mu=0,\sigma, \alpha = \{2, 1, 0, -2\}\right)$ where scale parameter $\sigma \sim \mathcal{H}\mathcal{N} (\gamma=5)$ and $\alpha$ is \{ST, AM, M, D\}, respectively.\\
    \midrule
    \textbf{player} &    &    & \checkmark  &  $\mathcal{S}\mathcal{N}\left(\mu=0,\sigma, \alpha = \{2,0\}\right)$ where scale parameter $\sigma \sim \mathcal{H}\mathcal{N} (\gamma=5)$ and $\alpha$ is assigned depending on prior beliefs about a player (\{2: good finisher, 0: not good finisher\}).\\
    \bottomrule
    \end{tabular}%
  \label{tab:priors}%
\end{table}%

Variables lacking a clear rationale for a positive or negative value are assigned a prior distribution modelled on the normal distribution. In contrast, variables, where the direction of the effect can be reasonably predicted, are assigned a prior distribution skewed towards that direction. This choice is made to favour values in the predicted direction while still allowing for the possibility of values in the opposite direction, acknowledging the potential for an incorrect prediction. The justifications for prior distribution choices are given below:
\begin{itemize}
    \item \textit{distance\_to\_goal} ($\mu=-1$ and $\alpha=-1$): Scoring becomes progressively more challenging as the distance from the goal increases, primarily because the goalkeeper gains additional time to react to the shot.

    \item \textit{shot\_angle} ($\mu=1$ and $\alpha=1$): This is because an increase in the shot angle implies either moving closer to the goal or positioning oneself more centrally, both of which result in a higher average scoring probability.

    \item \textit{players\_in\_shot\_triangle} ($\alpha=\left\{5,\dots,-5\right\}$): The presence of an increasing number of players within the shot triangle makes it increasingly challenging for a player to execute a shot in a manner that avoids hitting any of the players while still managing to score a goal.

    \item \textit{opponents\_in\_radius} ($\alpha=\left\{1,\dots,-2\right\}$): In a manner akin to the previously mentioned feature, an increase in the number of opponents within the shooter's proximity corresponds to an increased challenge for the shooter. With more players in the vicinity, there is an augmented effort from opponents to disrupt and block the shot, consequently making it more difficult for the shooter to successfully score.

    \item \textit{gk\_in\_shot\_triangle} ($\alpha=-2$): If the goalkeeper is positioned within the shot triangle, their chances of successfully saving a shot are higher compared to when they are located outside of the shot triangle.

    \item \textit{shot\_one\_on\_one} ($\alpha=2$): When a player finds themselves in a one-on-one situation with the goalkeeper, their sole task is to outplay the goalkeeper with their shot, without the need to navigate or consider other players. This circumstance makes scoring comparatively more straightforward.

    \item \textit{shot\_open\_goal} ($\alpha=4$): Similar to the previous feature, in this case, there is no goalkeeper present. Consequently, the shooter's sole objective is to direct the shot accurately towards the target, making the likelihood of scoring very high.

    \item \textit{under\_pressure} ($\alpha=-2$): When a player is under pressure, their ability to concentrate on accurate and well-targeted shooting diminishes, resulting in a decreased likelihood of scoring on average.

    \item \textit{general\_position} ($\alpha=\left\{ST:2, AM:1, M:0, D:-2\right\}$): The $\alpha$ values signify the expectation that strikers will exhibit the highest proficiency in finishing, followed by attacking midfielders, other midfielders, and, finally, a decrease for defenders.

    \item \textit{player} ($\alpha=\{2, 0\}$): If a player is anticipated to excel in finishing skills based on their name and reputation, they are given a value of 2 for $\alpha$; otherwise, a value of 0 is assigned.

    \item A standard value of $\sigma=5$ was chosen for the priors to strike a balance, ensuring sufficient variability. This choice aims to prevent the priors from becoming overly narrow in case the underlying prior knowledge is incorrect. Simultaneously, it avoids excessive largeness that could prolong convergence and necessitate numerous rounds of sampling.
\end{itemize}

\subsection{Model Development}
Following the exposition of technical aspects related to the models employed in this paper, we proceed to elucidate the practical details of their implementation. The entire computational implementation was carried out using the Python programming language, specifically version 3.8 and above. In non-Bayesian modelling phases, logistic regression was executed using the Python \textit{sklearn} module along with its associated methods.

Bayesian modelling stages were implemented by using the \textit{bambi} module which is an open-source Python package and purposefully crafted to simplify the fitting of generalised linear multilevel models (GLMMs) using a Bayesian framework. This encompasses a broad category of techniques widely employed in various research domains, including linear regression, analysis of variance (ANOVA), logistic and Poisson regression, as well as multilevel and crossed-group-specific effects models. Bambi facilitates the specification of intricate generalised linear hierarchical models through a formula notation reminiscent of R, providing a balance between user-friendly syntax for novices and direct access to internal objects for advanced users. This design allows beginners to swiftly articulate complex models with default priors, akin to popular R packages while providing seasoned users with the flexibility to directly manipulate internal objects for a more advanced and nuanced modelling approach \citep{Capretto2022}.

In particular, a Markov chain Monte Carlo model created in \textit{bambi} is used to produce posterior distributions with 1500 draws where 250 draws of which are chosen to be the burning period. In total, 4 Markov chain sampling is developed resulting in 6000 total samples for each \emph{Bayes-xG} model. We set a target acceptance ratio of 95\% in the model whilst using the prior distributions given in Table \ref{tab:priors}. Considering the target feature in the models is a binary variable of goal status, we decided to utilise a Bernoulli likelihood for all \emph{Bayes-xG} models. 

Furthermore, in the context of this paper, Bayesian implementation of the models was executed using data from a specific league. This decision was driven by the fact that a substantial portion of the entire dataset consisted primarily of Barcelona matches, as this information was publicly provided by StatsBomb. The concern was that including such a dominant dataset might introduce bias and influence the results. By focusing on a single league, the analysis can encompass a diverse range of players and team matchups, ensuring a more balanced and comprehensive examination.

\section{Experimental Analysis}
\label{sec:results}
The experimental analysis of this paper was studied under 5 cases:
\begin{enumerate}
    \item Frequentist model comparison to benchmark model for the whole 60K+ shots data set,
    \item Bayes-xG model-based ``\emph{positional}" analysis and comparisons for English Premier League data set (10K+ shots)
    \item Bayes-xG model-based ``\emph{player-specific}" analysis and comparisons for English Premier League data set (10K+ shots)
    \item Extending the developed Bayes-xG model evaluations into different countries, e.g. Spain (La Liga - 19K shots) and Germany (Bundesliga - 7.5K shots).
    \item Investigating the choice of priors on Bayes-xG model outputs
\end{enumerate}

\subsection{Frequentist / non-Bayesian models}
In the initial series of experiments, we assessed the modelling performance of frequentist models based on logistic regression, namely the Baseline xG and Extended xG, utilising two distinct sets of features. The analysis involved the complete data set comprising over 63,000 instances. The objective was to observe and compare how these models deviate from the predictions of the benchmark Statsbomb xG model. 

The distributions of the predicted xG values for each model are shown below in Figure \ref{fig:freq1}, along with the StatsBomb xG values for comparison. As expected, the Extended model performs better than the Baseline model (with shot angle, distance to goal, and the interaction between the two) by also predicting much more extreme values and by decreasing the interquartile and whisker ranges. Furthermore, the extended model has a distribution very close to that of the StatsBomb model.

\begin{figure}
    \centering
    \includegraphics[width=\linewidth]{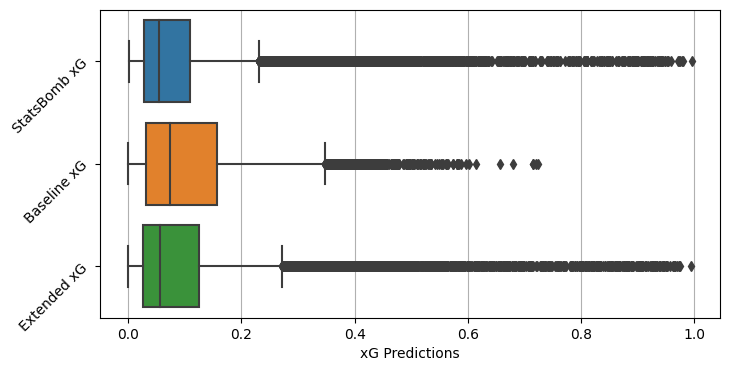}
    \caption{Distributions of Predictions from Frequentist xG Models}
    \label{fig:freq1}
\end{figure}

Regarding the assessment metrics, Table \ref{tab:fregPerf} illustrates the performance of each model in comparison to the Statsbomb xG model in terms of Brier score, $R^2$, mean absolute error (MAE) and root mean square error (RMSE). As anticipated, the extended model demonstrated superior performance over the Baseline model, given its enhanced information for predicting goal probability. Notably, the extended model exhibited a Brier score nearly identical to the StatsBomb model, indicating comparable performance to an industry-leading xG model according to this metric.

\begin{table}[htbp]
  \centering
  \caption{Outcomes from non-Bayesian models, including the Baseline xG model incorporating distance to the goal, shot angle, and their interaction, and the Extended xG model introducing additional features, in comparison to the StatsBomb xG model.}
    \begin{tabular}{p{3.75cm}p{3.75cm}p{3.75cm}p{3.75cm}}
    \toprule
    \multicolumn{1}{r}{} & Baseline xG & Extended xG & Statsbomb xG \\
    \midrule
    RMSE & 0.095 & 0.055 & \multicolumn{1}{p{4.055em}}{-} \\
    \midrule
    MAE & 0.058 & 0.029 & \multicolumn{1}{p{4.055em}}{-} \\
    \midrule
    $R^2$ Score & 0.428 & 0.826 & \multicolumn{1}{p{4.055em}}{-} \\
    \midrule
    Brier Score & 0.086 & 0.076 & 0.075 \\
    \bottomrule
    \end{tabular}%
  \label{tab:fregPerf}%
\end{table}%

\begin{figure}[t!]
    \centering
    \includegraphics[width=\linewidth]{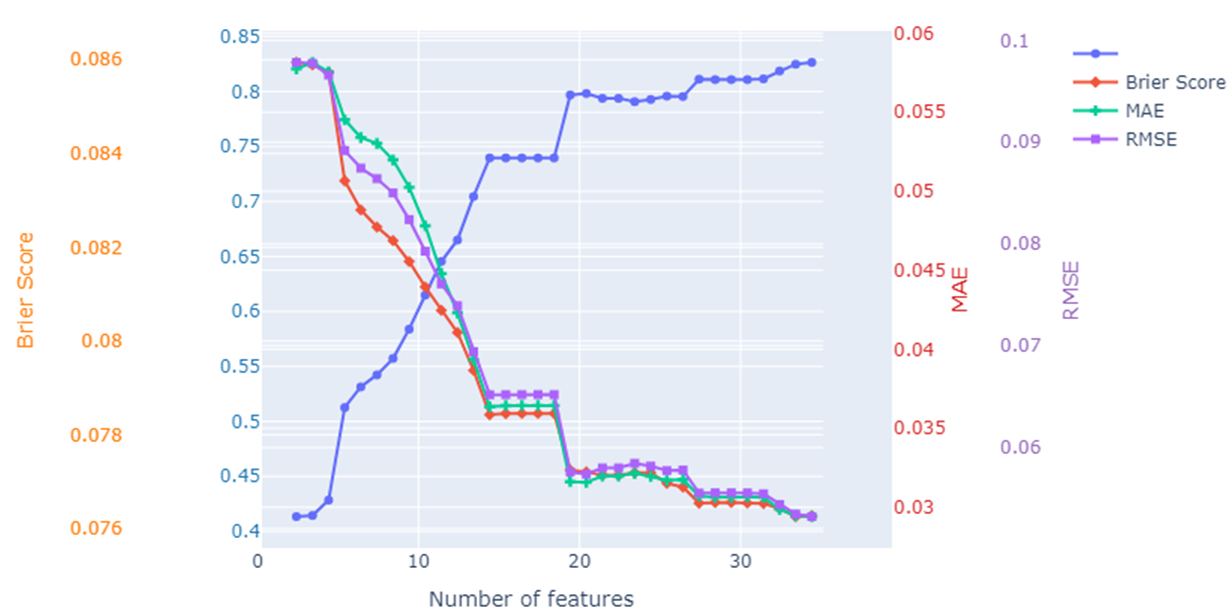}
    \caption{Model fitting performance when increasing the number of features.}
    \label{fig:freq2}
\end{figure}

In the final analysis of the initial set of experiments, we explored the impact of incorporating engineered advanced features on the performance of the frequentist logistic regression model. Figure \ref{fig:freq2} illustrates the trends in performance evaluation metrics—Brier score, R2, MAE, and RMSE—relative to the number of features integrated into the model. We initiated the analysis with a single-parameter model, utilizing only the distance to the goal, and systematically added features one by one. Typically, there are 16 model parameters (as detailed in Table \ref{tab:priors}), but this count increases to a maximum of 33 after one-hot encoding categorical features. Examining Figure \ref{fig:freq2} reveals a substantial influence on model performance resulting from the introduction of these created parameters. Notably, there are certain plateaus in the trends, particularly associated with one-hot encoded parameters representing categories with a minimal number of samples in the dataset (e.g., 10 players in the shot triangle).

\subsection{Positional analysis via Bayesian models}
For the second experimental case in this paper, we investigate the effects of the general player positions on the pitch to the xG values by performing a Bayesian hierarchical modelling approach. Baseline xG and Extended xG models are redeveloped by using the positional grouping effect and we obtained \emph{Bayes-xG}$_1$ and \emph{Bayes-xG}$_2$ models, respectively. 

We commence with a more straightforward model that examines only a few features within the logistic regression framework. As mentioned in the above sections, the baseline xG model incorporates distance to the goal, shot angle, and their interaction as predictors. To assess the influence of the grouping variable, "\emph{general position}" on xG, we calculate the differences in xG predictions between the single-level frequentist model and its hierarchical Bayesian counterpart, referred to as the \emph{Bayes-xG}$_1$ model, for each shot in the dataset.

\begin{figure}[t!]
    \centering
    \includegraphics[width=\linewidth]{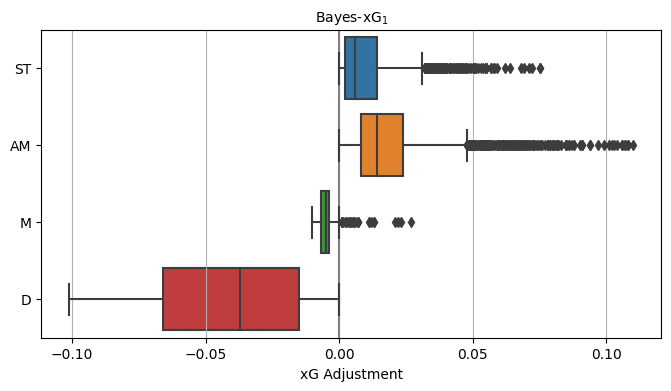}
    \caption{Distributions of xG Adjustments by Position of Bayes-xG$_1$.}
    \label{fig:model1a}
\end{figure}

Figure \ref{fig:model1a} illustrates the distributions of the xG adjustment, considering the general position within the hierarchical model for each position category. The observed distributions align well with the prior beliefs regarding the impact of the general position. Defenders exhibit a substantial number of negative xG adjustments in comparison to the baseline model's xG predictions, with some adjustments reaching as low as 0.1. As anticipated, non-attacking midfielders display smaller xG adjustments, encompassing both positive and negative values. Contrary to expectations, attacking midfielders exhibit larger positive xG adjustments on average compared to strikers, who also generally have positive adjustments. This unexpected finding may stem from the fact that strikers, by shooting more frequently from high xG scoring positions, often possess sufficiently high xG values without requiring a significant positional adjustment. On the other hand, attacking midfielders frequently take shots from areas around the goal, where xG chances are lower, yet they excel in scoring due to their above-average attacking and scoring capabilities.

\begin{figure}[t!]
    \centering
    \includegraphics[width=\linewidth]{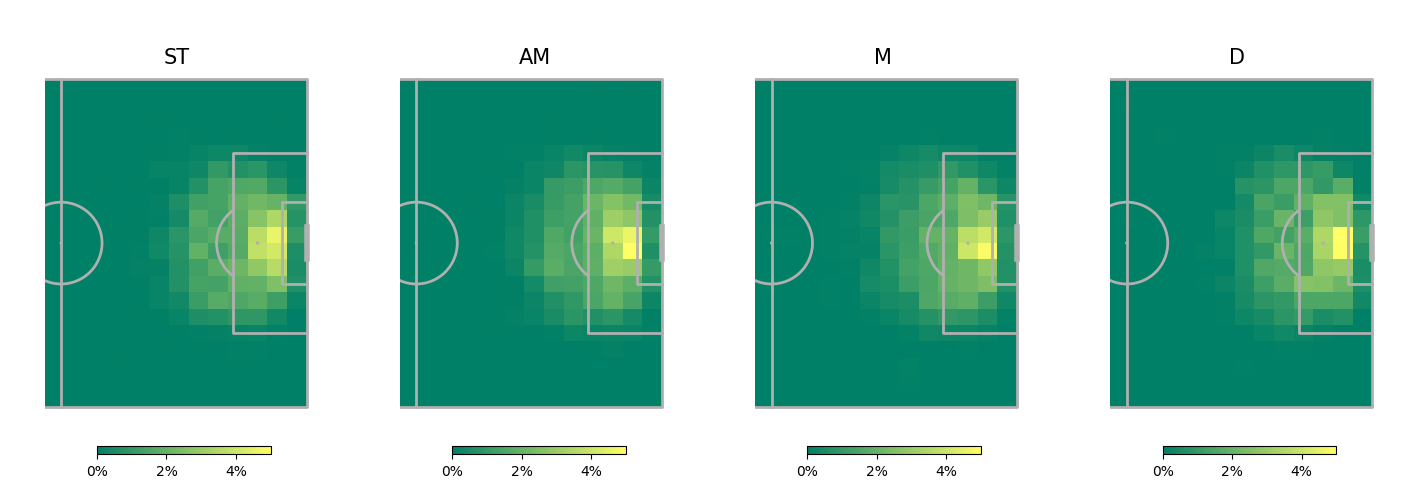}
    \caption{Normalized Heatmap of Shot Locations by General Position.}
    \label{fig:model1b}
\end{figure}

Figure \ref{fig:model1b} illustrates the shot locations categorised by general playing positions, normalised for each position. It highlights the notable concentration of chances for defenders positioned directly in front of the goal, providing a potential explanation for the observed reversals in Figure \ref{fig:model1a}. Contrary to the theorised expectation from the analysis in Figure \ref{fig:model1a}, attacking midfielders are not inclined to take shots from distant or challenging angles. In fact, on average, strikers exhibit a higher tendency for such shots. This discovery, coupled with the observation that attacking midfielders, on average, have larger positive xG adjustments than strikers, suggests that attacking midfielders may have a superior ability, on average, to convert high xG chances situated right in front of the goal compared to their striking counterparts.

Before delving into the more intricate model analysis for Bayes-xG$_2$, we aim to showcase a validation step to demonstrate the accuracy of the MCMC-based sampling technique employed in developing Bayesian models in this paper. To achieve this, we replicated the Bayesian analysis, this time utilising Bayes' Formula
\begin{align}
P(\text{goal}|\text{position}_i) = \dfrac{P(\text{position}_i|\text{goal}) \cdot P(\text{goal})}{P(\text{position}_i)}.
\end{align}
This allowed us to conduct an analysis where the results of the baseline model could be adjusted using Bayes’ Theorem, and these adjusted outcomes were then compared to the results of the hierarchical model. The comparison aimed to assess the proximity between theoretical adjustments and model adjustments. The outcomes of this process are detailed in Table \ref{tab:model1c}. Notably, the mean model adjustments closely align with the theoretical adjustments for each position, affirming that the model has effectively estimated the positional impact.

\begin{table}[t!]
  \centering
  \caption{Mean xG adjustment for each general position from Bayes-xG$_1$ versus theoretical adjustment of Baseline xG prediction using Bayes’ Theorem.}
    \begin{tabular}{p{3cm}p{6.5cm}p{6.5cm}}
    \toprule
    Position & Mean Model Adjustment & Mean Theoretical Adjustment \\
    \midrule
    ST & 0.009 & 0.010 \\
    \midrule
    AM & 0.019 & 0.020 \\
    \midrule
    M  & -0.006 & -0.005 \\
    \midrule
    D  & -0.042 & -0.044 \\
    \bottomrule
    \end{tabular}%
  \label{tab:model1c}%
\end{table}%

\begin{figure}[t!]
    \centering
    \includegraphics[width=\linewidth]{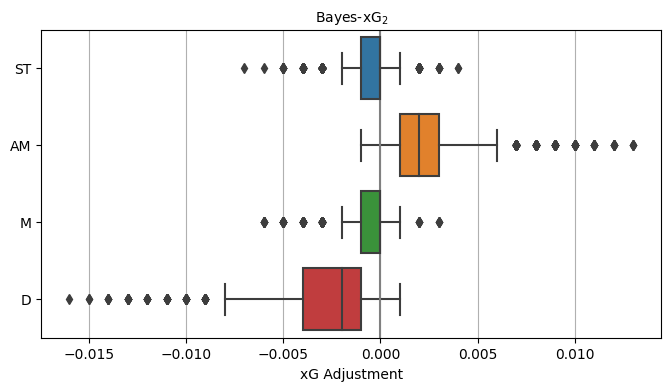}
    \caption{Distributions of xG adjustments for Bayes-xG$_2$, where adjustment is hierarchical model prediction minus baseline model prediction - grouped by general position}
    \label{fig:model2a}
\end{figure}

We proceed with our analysis by delving into the upgraded iteration of the Baseline model, referred to as the Extended model. Employing Bayes-xG$_2$, this advanced model involves a positional analysis akin to its predecessor, Bayes-xG$_1$. However, it incorporates numerous additional predictors, including factors such as \textit{opponents\_in\_radius} and \textit{gk\_distance\_to\_goal}, aiming to enhance the baseline xG predictions. Upon inspecting the distributions of xG adjustments by position in Figure \ref{fig:model2a}, it is evident that the values are notably smaller when compared to those derived from Bayes-xG$_1$. Few adjustments now exceed 0.01 away from the baseline xG values. This observation implies that the supplementary predictors effectively contribute to position prediction. This indicates that xG advantages stem less from the player's position and more from the specific play situation during shooting. For instance, while attackers generally enjoy better scoring chances on average, leading to positive xG adjustments in the basic model, the Extended model, by accounting for various features defining these improved scoring chances (e.g., one-on-ones), mitigates the impact of player position on xG adjustments.

\begin{figure}[t!]
    \centering
    \includegraphics[width=\linewidth]{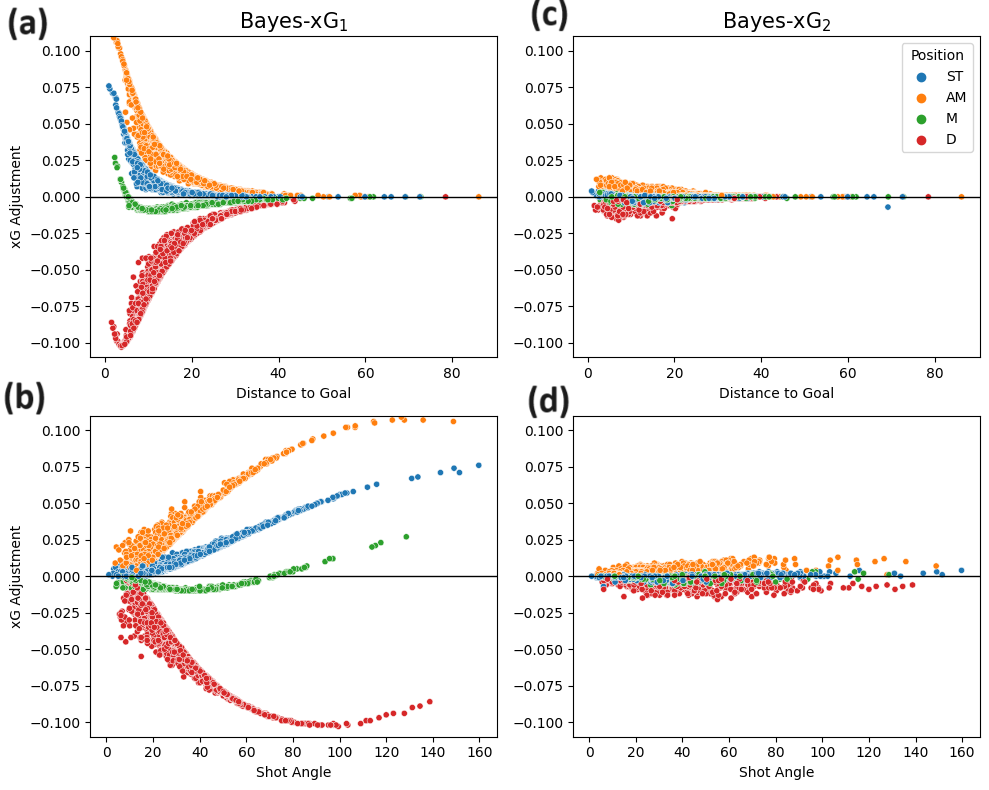}
    \caption{Comparison of point estimates for xG adjustments against \textit{distance\_to\_goal} and \textit{shot\_angle} between Bayes-xG$_1$ and Bayes-xG$_2$, grouped by general position. Adjustments are hierarchical model prediction minus baseline model prediction.}
    \label{fig:model2b}
\end{figure}

It is intriguing to examine the xG adjustments across the predictor ranges of "\textit{distance\_to\_goal}" and "\textit{shot\_angle}." Figure \ref{fig:model2b} illustrates these xG adjustments grouped by the position for both Bayes-xG models. In Figure \ref{fig:model2b}-(a) for Bayes-xG$_1$, the convergence of each position towards an adjustment of 0 signifies that, at a certain distance, the xG value tends to be close to zero, regardless of other shot-related factors. Notably, defenders exhibit a slight dip in xG adjustments at the lowest distance before increasing. This suggests scenarios where defenders find themselves in goal-scoring positions, perhaps following set pieces or during a team's pursuit in a match. The order of adjustments in Figure \ref{fig:model1a} aligns consistently across positions, with attacking midfielders having the highest positive adjustments, followed by strikers, and other midfielders showing minimal adjustment from the baseline model. Examining the "\textit{shot\_angle}" for Bayes-xG$_1$ in Figure \ref{fig:model2b}-(b), smaller values correspond to more challenging scoring opportunities, either due to being far from the goal or from tight angles. Defenders display gradually larger negative adjustments as the shot angle increases, reversing after a certain point for very high shot angles, likely corresponding to very close distances to the goal. Similar patterns emerge for other positions, with attacking midfielders having the largest positive xG adjustments, followed by strikers and other midfielders. For Bayes-xG$_2$ model outputs in Figure \ref{fig:model2b}-(c) and (d), the compensation of positional effects through additional predictors is evident. Despite a clear distinction in the effects of distance to the goal and shot angle for each position in Bayes-xG$_1$, no significant differences between positions are observed in Bayes-xG$_2$. The diminishing impact of position on xG adjustment is attributed to the diverse player abilities and roles within each position category. A distinct observation evident in both Figure \ref{fig:model2b}-(c) and (d) is that defenders exhibit consistent trends for all angles just below 0, whereas attacking midfielders mirror the same pattern but above the 0 line. Adjustments made to strikers' and midfielders' xG values are barely discernible, with values hovering around 0 across all shot angles.  

To further explore the aforementioned phenomena, the next experimental case involves a player-specific analysis with Bayes-xG$_3$, grouping data based on the player taking the shot rather than their general position.

\subsection{Player-specific analysis via Bayesian models}
In the context of the third experimental case explored in this paper, we introduced Bayes-xG$_3$, derived from the Extended model, with players designated as the grouping variable in Bayesian hierarchical modelling. The focus here is to assess whether the models necessitate player-specific adjustments. Unlike the previous experimental case that centred on positional analysis, conducting a player-specific analysis poses increased complexity due to the considerably larger pool of candidates within the group, making the analysis more challenging and computationally intensive. Instead of individually representing each player in our group, a selective approach is employed, categorising the majority as "other" and opting for a few players with expected positive or negative xG adjustments. Player selection is based on the "conversion rate," i.e., the percentage of shots scored. To ensure relevance, only players with a minimum of 50 shots are considered, and the chosen players, along with their statistics, are detailed in Table \ref{tab:model3a}. Players like R. Pirès, S. Agüero, and J. Vardy, recognized for their prolific goal-scoring, are expected to have positive xG adjustments. Pirès, notably, exhibits an exceptional conversion rate in the data subset. Conversely, players like P. Coutinho and R. Barkley, with below-average conversion rates, might have slight negative xG adjustments, while J. Shelvey, who failed to convert any of his 51 shots in the data, is likely to have a more substantial negative xG adjustment.

\begin{table}[t!]
  \centering
  \caption{Selected players for Bayes-xG$_3$ and their goal-scoring statistics in the data set.}
    \begin{tabular}{p{5cm}p{3cm}p{3cm}p{4cm}}
    \toprule
    Player & Shots & Goals & Conversion Rate \\
    \midrule
    Robert Pirès & 56 & 14 & 25.00\% \\
    \midrule
    Sergio Agüero & 112 & 20 & 17.90\% \\
    \midrule
    Jamie Vardy & 111 & 19 & 17.10\% \\
    \midrule
    Phillippe Coutinho & 105 & 8  & 7.60\% \\
    \midrule
    Ross Barkley & 82 & 6  & 7.30\% \\
    \midrule
    Jonjo Shelvey & 51 & 0  & 0.00\% \\
    \bottomrule
    \end{tabular}%
  \label{tab:model3a}%
\end{table}%

As explained in the preceding methodology section, the impact of each player will be characterised by a prior distribution, specifically a skewed normal distribution. The choice of distribution parameters is dependent upon the prior beliefs regarding a player's proficiency as a goal scorer, with the parameter $\alpha$ taking values of either 2 or 0. This selection is guided by qualitative beliefs about the players rather than direct utilisation of the data for informing the priors. Players such as Pirès, Agüero, Vardy, and Coutinho, acknowledged as talented attacking players, are attributed $\alpha=2$. In contrast, Barkley and Shelvey, who are not commonly associated with being top-tier attackers but possess other defining qualities in their game, are assigned $\alpha=0$.

\begin{figure}[t!]
    \centering
    \includegraphics[width=\linewidth]{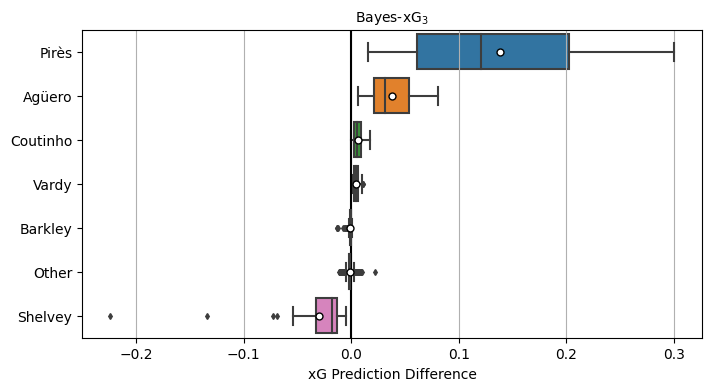}
    \caption{Distributions of xG adjustments for Bayes-xG$_3$, where adjustment is hierarchical model prediction minus baseline model prediction - grouped by player.}
    \label{fig:model3b}
\end{figure}

Figure \ref{fig:model3b} illustrates the distributions of xG adjustments for individual players and the collective "other" players group. A prominent observation is the substantial positive xG adjustments for Robert Pirès, some reaching as high as 0.3 above the baseline xG. Notably, these adjustments persist even after incorporating additional predictors in Bayes-xG$_2$ that were intended to eliminate group effects in the previous experimental analysis. Pirès also exhibits a wide spread of adjustments, ranging close to 0, indicating a diverse array of shot types. Some were high xG chances, requiring minimal adjustment, while others were more challenging but consistently converted by Pirès, resulting in significant positive adjustments. Agüero displays consistently positive xG adjustments, albeit smaller on average and with a narrower spread compared to Pirès. Intriguingly, Vardy and Coutinho exhibit minimal positive xG adjustments, not significantly greater than those of Barkley. Shelvey aligns with expectations, displaying substantial negative xG adjustments based on his conversion rate in the data. Lastly, the "other" group centres around 0 for xG adjustment, as anticipated, given its diverse player composition with no discernible group effect to capture.

\begin{figure}[t!]
    \centering
    \includegraphics[width=\linewidth]{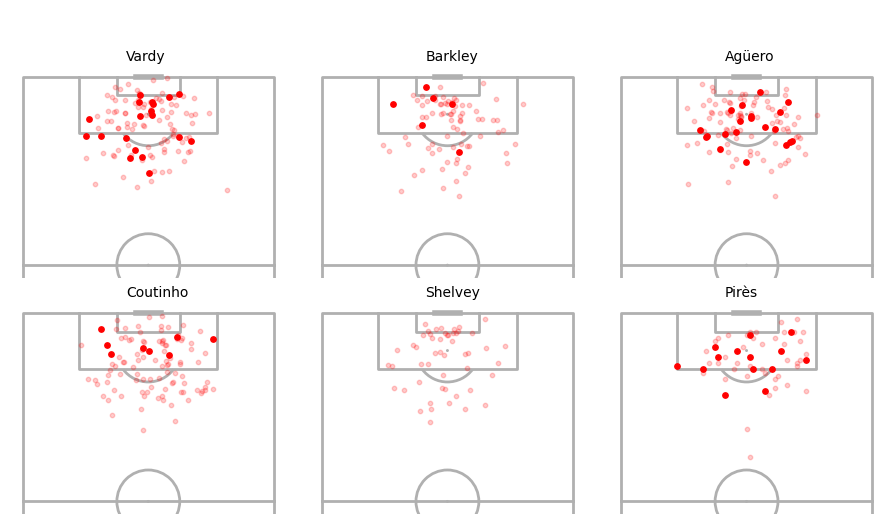}
    \caption{Selected player shot locations and goals.}
    \label{fig:model3c}
\end{figure}

Figure \ref{fig:model3c} displays the shot locations and outcomes for the selected players, offering insights into the findings presented in Figure \ref{fig:model3b}. Beginning with Pirès, notable for his substantial positive xG adjustments, the observation centres on his efficiency in goal scoring despite a relatively low number of shots. His ability to score from challenging positions, such as both corners of the box and outside the area, contributes to the positive xG adjustments, indicating his prowess as a goal scorer even in demanding scenarios. Agüero and Vardy exhibit similar shot patterns, but the model assigns significantly higher positive xG adjustments to Agüero. This discrepancy may stem from the nature of Vardy's shots being inherently high xG chances, like one-on-one opportunities, whereas Agüero manages to convert more challenging shots, resulting in larger adjustments. Comparing Vardy with Coutinho and Barkley, who exhibit similar xG adjustments in Figure \ref{fig:model3b}, suggests that their goal-scoring patterns align with baseline xG values without substantial player adjustments. Lastly, Shelvey's shot map lacks goals from various positions. While difficult-to-score shots receive minor adjustments, centrally located missed chances likely contribute to the notable negative adjustments, reflecting Shelvey's poor conversion rates in this dataset.

\begin{figure}[t!]
    \centering
    \includegraphics[width=\linewidth]{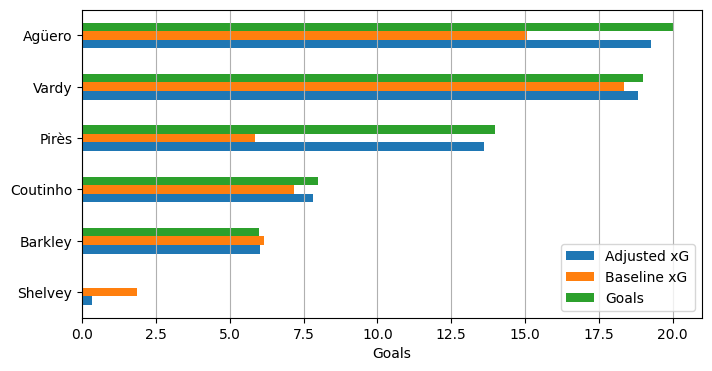}
    \caption{Comparison of Baseline, Bayes-xG$_3$ hierarchical predictions, and actual goals scored for selected players.}
    \label{fig:model3d}
\end{figure}

Figure \ref{fig:model3d} displays the cumulative data for goals scored, baseline expected goals (xG) from the single-level model, and adjusted xG from the player-corrected model for the selected players in this analysis. The visual representation illustrates that, in comparison to the single-level model, the player-corrected model provides more accurate estimates of total goals scored by each player. Notably, players like Pirès and Agüero, who outperformed their baseline xG by scoring difficult chances, exhibit adjusted xG totals much closer to their actual goals scored. Conversely, Shelvey's adjusted xG total is more aligned with the zero goals he scored, although it is crucial to emphasise that it is not precisely zero.

\subsection{Extension into other leagues}
To ascertain the generalizability of the conclusions drawn in this study beyond the Premier League dataset examined earlier and their applicability to football on a broader scale, as our fourth experimental case, a parallel analysis was conducted using data from Spain's La Liga and Germany's Bundesliga. The datasets for these leagues encompass approximately 19,000 and 7,500 shots, respectively. It is noteworthy that the La Liga dataset is notably influenced by Barcelona, primarily due to the fact that StatsBomb predominantly released data from games in which Lionel Messi, a prominent Barcelona player, participated.

\begin{figure}[t!]
    \centering
    \includegraphics[width=\linewidth]{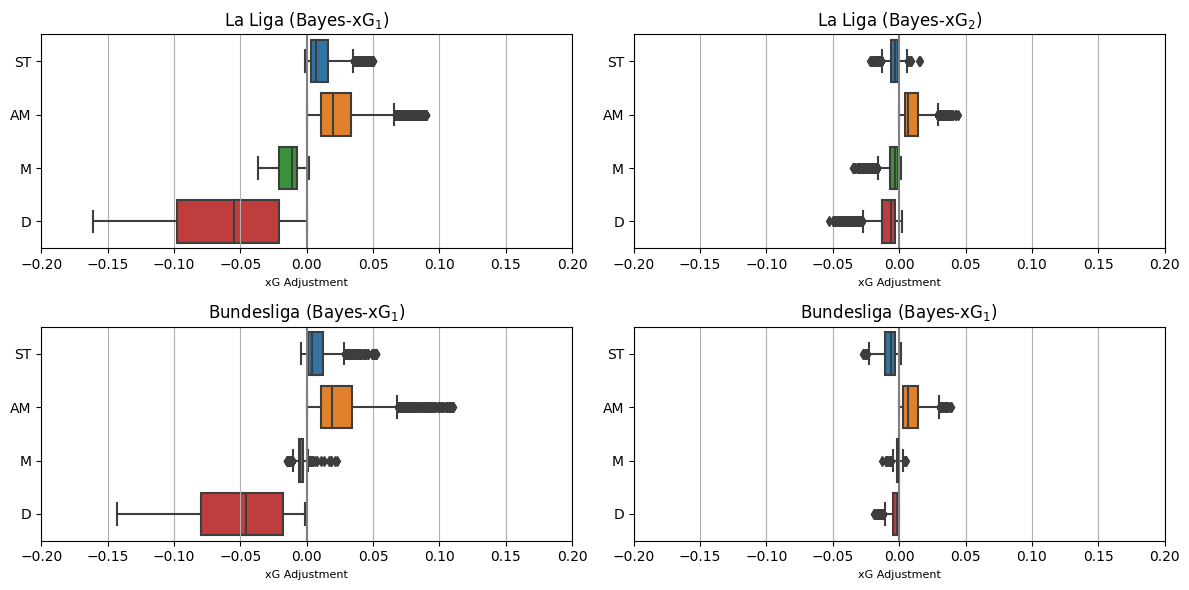}
    \caption{Distributions of xG adjustments for Bayes-xG$_1$ and Bayes-xG$_2$ for Spanish La-Liga and the German Bundesliga.}
    \label{fig:model4a}
\end{figure}

Figure \ref{fig:model4a} displays the distributions of xG adjustments for the Bundesliga and La Liga, which serve as the data sets for the Bayes-xG$_1$ and Bayes-xG$_2$ models. The outcomes for both leagues closely resemble those of the Premier League in the Bayes-xG$_1$ model. The average adjustments follow a consistent order across all leagues, with attacking midfielders exhibiting the most positive adjustments, followed by strikers, other midfielders, and then a substantial drop to defenders. Additionally, the magnitudes of these adjustments exhibit similar patterns. The Bayes-xG$_2$ model reveals a noteworthy reduction in the magnitude of xG adjustments, a trend observed consistently across all leagues. Although the Bayes-xG$_2$ model indicates slightly larger adjustment magnitudes for the additional leagues, the variation is not significant enough to alter the model results significantly when compared to those of the Premier League.

To conduct a player-specific examination of Bayes-xG$_3$, the inclusion of new players from both leagues was necessary. The selection process mirrored that of the Premier League, where players were listed based on their conversion rates in the data sets, encompassing both the best and worst performers. Tables \ref{tab:model4b} and \ref{tab:model4c} display the chosen players and their respective statistics for La Liga and the Bundesliga, providing a comprehensive overview of the selected players' performance in each league.

\begin{table}[t!]
  \centering
  \caption{Selected players for Bayes-xG3 and their scoring statistics from La Liga data set.}
    \begin{tabular}{p{5.5cm}p{3cm}p{3cm}p{4cm}}
    \toprule
    Player & Shots & Goals & Conversion Rate \\
    \midrule
    Gareth Bale & 89 & 20 & 22.50\% \\
    \midrule
    Lionel Messi & 1862 & 375 & 20.10\% \\
    \midrule
    Samuel Eto’o & 295 & 62 & 21\% \\
    \midrule
    Bebé & 74 & 2  & 2.70\% \\
    \midrule
    Rafael Márquez & 53 & 2  & 3.80\% \\
    \midrule
    Andrés Iniesta & 362 & 25 & 6.90\% \\
    \bottomrule
    \end{tabular}%
  \label{tab:model4b}%
\vspace{0.25cm}
  \centering
  \caption{Selected players for Bayes-xG3 and their scoring statistics from Bundesliga data set.}
    \begin{tabular}{p{5.5cm}p{3cm}p{3cm}p{4cm}}
    \toprule
    Player & Shots & Goals & Conversion Rate \\
    \midrule
    Javier Hernández & 63 & 16 & 25.40\% \\
    \midrule
    Pierre-Emerick Aubameyang & 107 & 22 & 20.60\% \\
    \midrule
    Robert Lewandowski & 147 & 28 & 19\% \\
    \midrule
    Pascal Groß & 56 & 1  & 1.80\% \\
    \midrule
    Hakan Çalhanoğlu & 51 & 1  & 2\% \\
    \midrule
    Timo Werner & 64 & 6  & 9.40\% \\
    \bottomrule
    \end{tabular}%
  \label{tab:model4c}%
\end{table}%

\begin{figure}[t!]
    \centering
    \includegraphics[width=\linewidth]{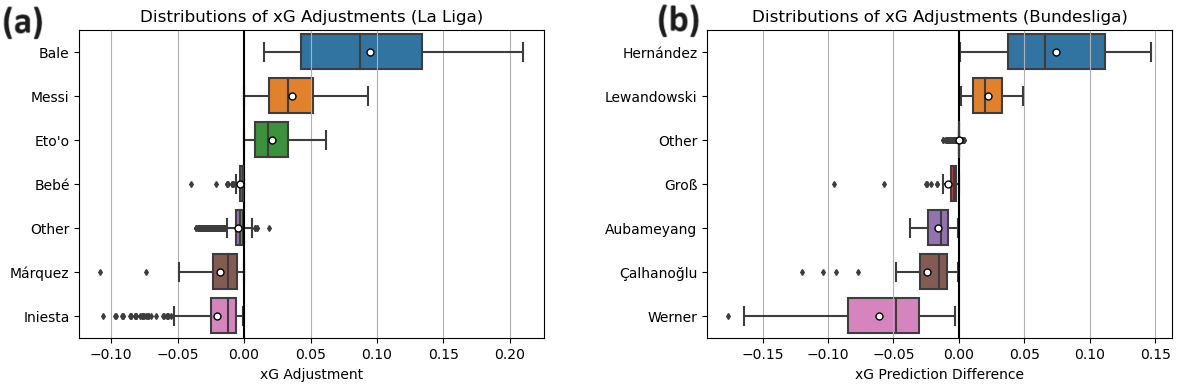}
    \caption{Distributions of xG adjustments for Bayes-xG$_3$ for Spanish La-Lida and the German Bundesliga}
    \label{fig:model4d}
\end{figure}

Given these conversion rates, it is reasonable to infer that the top three performers in each table would, on average, experience positive adjustments in expected goals (xG), while the bottom three would, on average, encounter negative or negligible xG adjustments. The visual representation in Figure \ref{fig:model4d} illustrates these adjustments for players in both the Spanish La Liga (a) and the German Bundesliga (b). As anticipated, players like Bale, Messi, and Eto'o from La Liga predominantly exhibit positive xG adjustments, aligning with expectations. Conversely, other selected players either display minimal adjustments or predominantly negative adjustments. In the Bundesliga context, Figure \ref{fig:model4d}-(b) reaffirms many anticipated outcomes. An intriguing finding is that Aubameyang, despite maintaining a high conversion rate, exhibits predominantly negative xG adjustments. This suggests that the goals he scores tend to be from chances with already high expected goals, contrary to some expectations.

\begin{figure}[t!]
    \centering
    \includegraphics[width=\linewidth]{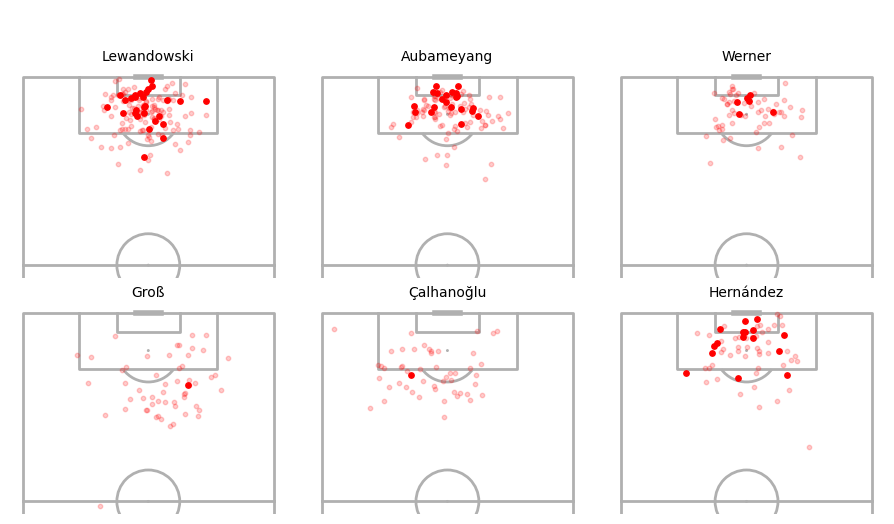}
    \caption{Selected player shot locations and goals for the German Bundesliga}
    \label{fig:model4e}
\end{figure}

Figure \ref{fig:model4e} illustrates the shot locations of Bundesliga players, validating that the majority of Aubameyang's shots originate from within the penalty area. This observation implies that these shots likely possess additional characteristics, such as one-on-one opportunities, making them high expected goals (xG) chances. In contrast, Timo Werner exhibits minimal goals despite a comparable shot map, and the substantial negative xG adjustments suggest that he should have scored more from these shot positions. On a different note, Çalhanoğlu records relatively few goals from shots that present higher difficulty due to their distance from the goal, resulting in slightly fewer negative xG adjustments.

In conclusion, the findings from both La Liga and the Bundesliga verify the results obtained in the Premier League sections above. Notably, there is an indication of a positional impact on xG in a fundamental xG model (Bayes-xG$_1$), but such effects markedly diminish with the adoption of a more intricate model (Bayes-xG$_2$). Nevertheless, even with the extended model, there remains evidence of player-specific effects on xG, providing a quantitative measure of how certain players excel or lag behind others in scoring.

\subsection{Analysing the choice of priors}
The impact of prior choices is particularly evident when assessing the efficiency and accuracy of Bayesian models. Sensitivity to the choice of priors is crucial, and careful consideration is needed to strike a balance between the informativeness of priors and their impact on computational efficiency. In Bayesian hierarchical modelling, this emphasizes the need for a thoughtful approach to prior selection, ensuring that the chosen priors align with the characteristics of the data and contribute to the model's robustness and reliability.

The selection of prior distributions holds significant importance in Bayesian hierarchical modelling for several reasons. Firstly, the choice of prior widths plays a pivotal role. Employing overly wide priors, encompassing a large range of values, may necessitate an extensive number of samples for the model to converge to the true values of the variables, leading to prolonged computation times. Conversely, if the prior distributions are excessively narrow and the true values are unlikely to be sampled, the model's outcomes may be biased and inaccurate.

To evaluate the appropriateness of the chosen prior distributions, a reassessment will be conducted by refitting the extended baseline model (single-level) with different priors, followed by a comparison of predictions. The sets of priors under consideration include (1) the existing priors given in Table \ref{tab:priors}, (2-3) a pair of wide-tight uniform priors, (4-5) a pair of wide-tight normal priors, and (6) a deliberately ill-suited set of priors. The choices of priors for the cases of (2-6) are presented in Table \ref{tab:priors2}. For the uniform priors, each variable has been given a wide and tight uniform distribution pair. Similarly, for normal priors, zero mean priors are chosen with two different $\sigma$ values to represent a wise and tight value support. On the other hand, the ill-suited priors have been given very narrow distributions by using a small value for $\sigma$. Moreover, some of the skews in the distributions have been flipped such that the prior belief about the effect is the reverse of what was actually used.

Furthermore, the predictions generated by these models will be juxtaposed with those of the extended non-Bayesian model on the same data, providing a baseline for assessing the efficacy of the selected priors in yielding accurate predictions.

\begin{table}[t!]
  \centering
  \caption{Choices of prior distributions for analysing the impact of prior choice on results.}
    \begin{tabular}{p{4.1cm}|p{1.5cm}|p{1.5cm}|p{1.5cm}|p{1.5cm}|p{4.25cm}}
    \toprule
     & 2-Wide & 3-Tight & 4-Wide & 5-Tight & 6-Ill-suited \\
    Predictor & Uniform & Uniform & Normal & Normal & Prior \\
    \midrule
    \textit{Intercept} & \multirow{14}[28]{*}{\parbox[t]{2mm}{\multirow{3}{*}{\rotatebox[origin=c]{60}{$\mathcal{U}(-100, 100)$}}}} & \multirow{14}[28]{*}{\parbox[t]{2mm}{\multirow{3}{*}{\rotatebox[origin=c]{60}{$\mathcal{U}(-1, 1)$}}}} & \multirow{14}[28]{*}{\parbox[t]{2mm}{\multirow{3}{*}{\rotatebox[origin=c]{60}{$\mathcal{N}(0, 10)$}}}} & \multirow{14}[28]{*}{\parbox[t]{2mm}{\multirow{3}{*}{\rotatebox[origin=c]{60}{$\mathcal{N}(0, 0.25)$}}}} &  $\mathcal{S}\mathcal{N}\left(0,0.25,2\right)$\\
\cmidrule{1-1}\cmidrule{6-6}    \textit{distance\_to\_goal} &    &    &    &    &  $\mathcal{S}\mathcal{N}\left(0,0.25,2\right)$\\
\cmidrule{1-1}\cmidrule{6-6}    \textit{shot\_angle} &    &    &    &    &  $\mathcal{N}\left(0,0.25\right)$\\
\cmidrule{1-1}\cmidrule{6-6}    \textit{distance*angle} &    &    &    &    &  $\mathcal{S}\mathcal{N}\left(0,0.25,-2\right)$\\
\cmidrule{1-1}\cmidrule{6-6}    \textit{gk\_distance\_to\_goal} &    &    &    &    &  $\mathcal{N}\left(0,0.25\right)$\\
\cmidrule{1-1}\cmidrule{6-6}    \textit{players\_in\_shot\_triangle} &    &    &    &    &  $\mathcal{S}\mathcal{N}\left(0,0.25,\{-5, \dots,5\}\right)$\\
\cmidrule{1-1}\cmidrule{6-6}    \textit{opponents\_in\_radius} &    &    &    &    &  $\mathcal{S}\mathcal{N}\left(0,0.25,\{1,\dots,-2\}\right)$\\
\cmidrule{1-1}\cmidrule{6-6}    \textit{shot\_body\_part} &    &    &    &    &  $\mathcal{N}\left(0,0.25\right)$\\
\cmidrule{1-1}\cmidrule{6-6}    \textit{shot\_first\_time} &    &    &    &    &  $\mathcal{N}\left(0,0.25\right)$\\
\cmidrule{1-1}\cmidrule{6-6}    \textit{gk\_in\_shot\_triangle} &    &    &    &    &  $\mathcal{S}\mathcal{N}\left(0,0.25,2\right)$\\
\cmidrule{1-1}\cmidrule{6-6}    \textit{shot\_one\_on\_one} &    &    &    &    &  $\mathcal{S}\mathcal{N}\left(0,0.25,-2\right)$\\
\cmidrule{1-1}\cmidrule{6-6}    \textit{shot\_open\_goal} &    &    &    &    &  $\mathcal{S}\mathcal{N}\left(0,0.25,-4\right)$\\
\cmidrule{1-1}\cmidrule{6-6}    \textit{shot\_technique} &    &    &    &    &  $\mathcal{N}\left(0,0.25\right)$\\
\cmidrule{1-1}\cmidrule{6-6}    \textit{under\_pressure} &    &    &    &    &  $\mathcal{S}\mathcal{N}\left(0,0.25,2\right)$\\
    \bottomrule
    \end{tabular}%
  \label{tab:priors2}%
\end{table}%

\begin{figure}[t!]
    \centering
    \includegraphics[width=\linewidth]{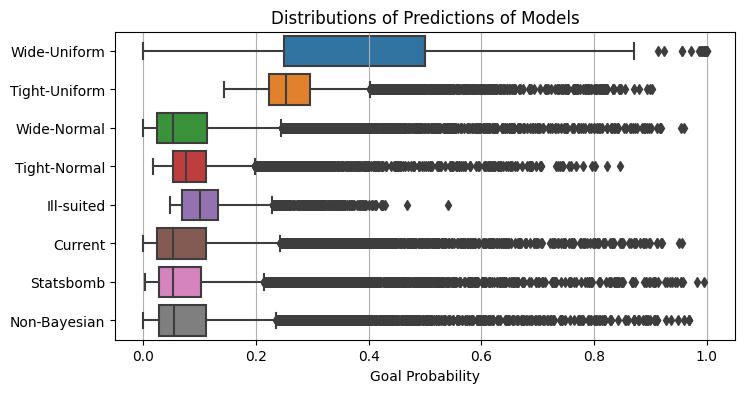}
    \caption{Distributions of xG predictions for each of the extended single-level Bayesian model, with different choices of prior distributions.}
    \label{fig:5a}
\end{figure}

Figure \ref{fig:5a} illustrates the distributions of predictions generated by each of the prior models. The model utilising wide uniform prior distributions exhibits notably poor performance when compared to both the non-Bayesian baseline model and the Statsbomb benchmark, displaying a considerable spread of predictions. On the other hand, employing tight uniform priors leads to more restricted predictions, although the average expected goals (xG) predictions tend to be relatively smaller. The utilisation of a tight normal prior yields similar performance, primarily underestimating xG values, particularly with the highest xG values concentrated around 0.8. In contrast, adopting wide normal priors results in enhanced performance compared to the tight setting, with predictions following a similar trend to the existing prior configuration. It is noteworthy that the mean and interquartile ranges of both normal priors depicted in Figure \ref{fig:5a} exhibit a favourable correspondence with the baseline and benchmark models.

On the other hand, as depicted in Figure \ref{fig:5a}, ill-suited priors result in a notably narrow spread, with scarce xG predictions exceeding 0.5. Despite this, the ill-suited priors exhibit improved performance compared to uniform priors in terms of aligning the average with the baseline and benchmark predictions and maintaining a similar-sized interquartile range. This improved performance is likely attributed to the greater number of samples utilised for parameter estimation. In the case of the model with uniform priors, the same number of samples, however, prevented the model from converging to optimal parameter values, resulting in poor predictions. While, given more samples, this model could eventually yield accurate results, the computational time required is uncertain and could be extensive. The model with the existing prior distributions outperforms the others significantly, closely resembling the distribution of baseline and benchmark models' predictions while also offering valuable insights based on player positions.

\begin{figure}[t!]
    \centering
    \includegraphics[width=\linewidth]{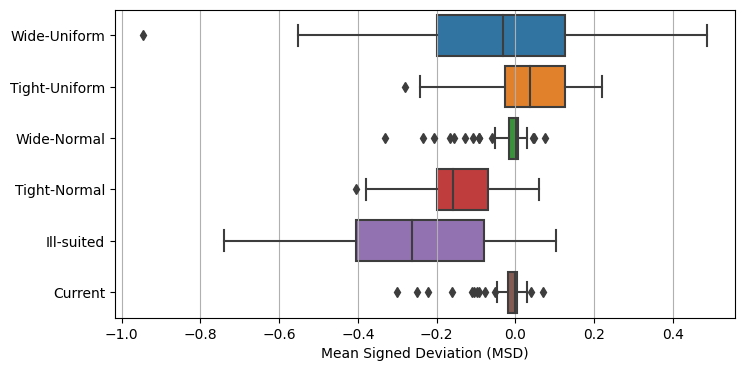}
    \caption{Mean signed deviation distributions for each analysed prior choice.}
    \label{fig:5b}
\end{figure}

In this section's final analysis, we explore the mean signed deviation (MSD) values between each prior case and the predictions of the Non-Bayesian extended model. The selection of MSD aims to emphasize instances of over or underprediction based on the prior choice, using the mean spread of MSD values as a performance metric. Figure \ref{fig:5b} illustrates the distributions of MSD values for each prior group through various box plots. Notably, Figure \ref{fig:5b} highlights the considerable performance of the Wide Normal prior choice, exhibiting results akin to the current prior approaches. Its interquartile range closely aligns with the current prior case, albeit with a few more instances of over-predicted outliers. It is crucial to observe that both uniform priors' MSD values are distributed around zero, despite with a broader spread. In contrast, the Tight-Normal and Ill-suited prior cases exhibit notably poor performance, marked by a higher frequency of overestimated xG values compared to the Non-Bayesian extended model. 

\section{Final Remarks \& Conclusion}
The objective of this study was to explore whether there exist distinct effects based on player position on expected goal (xG) values, suggesting that different positions or players may exhibit specific xG adjustments for given scoring opportunities. The hypothesis was that proficient attacking players, such as strikers, attacking midfielders, or those recognised for their offensive prowess, would demonstrate positive xG adjustments. In contrast, players less renowned for their attacking contributions, such as defenders or those with defensive roles, were expected to show negative xG adjustments.

To reach the objective mentioned above, this study has developed several Bayesian models to evaluate the influence of a player's position and individual player effects on xG predictions. Initially, a basic xG model (Baseline xG), incorporating only distance to the goal, shot angle, and their interaction, indicated positional effects on xG (Bayes-xG$_1$). Strikers and attacking midfielders exhibited positive xG adjustments, midfielders displayed minimal adjustments, while defenders had notably negative xG adjustments on average. However, the introduction of additional predictors in the models diminished the positional effects to the extent that they became almost insignificant (Extended xG), suggesting that player position had minimal impact on xG when considering more shot-related factors (Bayes-xG$_2$). Subsequently, player effects were explored using the extended model employed for the second positional-effects model (Bayes-xG$_3$), grouping the data based on the player's shooting rather than the shooter's position. The model was illustrated using six players from each dataset from three of the European Top 5 leagues, revealing significant player effects on xG even when controlling for various shot-related factors. These effects were diverse in direction, notably positive for R. Pirés (as well as for G. Bale and J. Hernandez) and negative for J. Shelvey (as well as for A. Iniesta and T. Werner).

The indication that there exist player-specific effects in determining goal probability could prove beneficial in football scouting and player selection. By computing adjusted xG values for various players and comparing these adjusted values to their non-adjusted counterparts, as demonstrated in this analysis, it becomes possible to distinguish players who excel at converting challenging opportunities from those who consistently find themselves in advantageous positions. Examining the results for the English Premier League dataset, particularly for J. Vardy, reveals that his total adjusted xG is not significantly different from his baseline xG (see Figure \ref{fig:model3d}). This suggests that, given the quality of chances Vardy receives, he scores at a relatively average rate. On the other hand, S. Agüero demonstrates a more consistent ability to score from more challenging positions, evident in his larger adjusted xG. It is important to note that this observation does not imply that Agüero is a superior player or attacker compared to Vardy. Instead, it suggests that, on average, Agüero is more adept at converting chances with lower xG values than Vardy, indicating proficiency in scoring from less favourable situations.

It is important to acknowledge that there might also be team-related influences at play in this context. To further compare Vardy and Agüero, Vardy is part of a Leicester team known for its high-tempo and direct attacking style. This approach likely leads to shooting scenarios where the ball is played behind the defence, creating situations with fewer defenders to obstruct or impede a shot. This dynamic often results in one-on-one opportunities with the goalkeeper, contributing to Vardy consistently receiving numerous high xG chances. Conversely, Agüero played for one of the top teams in the league, causing opponents to adopt a more conservative approach. Teams facing Manchester City tend to minimise space, making chances more challenging with multiple players in the shot triangle and other complicating factors.

The Bayesian modelling results were collocated with non-Bayesian, or frequentist, modelling. Particularly in the case of player correction, Bayesian modelling offers a significant advantage by capturing uncertainty through posterior distributions rather than relying solely on point estimates. This becomes particularly advantageous when dealing with younger players with limited match experience, as Bayesian hierarchical modelling effectively addresses data groups with few observations. Regardless, Bayesian modelling is rarely used in the literature of football analytics. Beyond its applications in scouting and player selection, Bayesian hierarchical modelling holds promise for various metrics in football. For instance, the assessment of injury risk among players, is a common practice in large football clubs, where certain players may have a higher overall susceptibility to injuries. Hierarchical modelling, in this context, has the potential to provide more accurate assessments of individual players' injury risks by considering their specific injury history.

\bibliography{main}

\end{document}